\documentclass[10pt, draft, showpacs, amsmath, amssymb, prd, twocolumn]{revtex4}
\def\a{\alpha}
\def\b{\beta}
\def\g{\gamma}
\def\d{\delta}

\def\vt{\vartheta}

\def\A{{\cal{A}}}

\def\a{\alpha}
\def\b{\beta}

\def\g{\gamma}
\def\d{\delta}
\def\g{\gamma}

\def\T{{\mathcal{T}}}

\def\F{{\mathcal{F}}}

\def\G{\Gamma}
\def\oG{\stackrel{{\rm o}}{\Gamma}{\!}}
 \def\oT{\stackrel{{\rm o}}{T}{\!}}
\def\LG{\stackrel{{\rm *}}{\Gamma}{\!}}
\def\oT{\stackrel{{\rm o}}{T}{\!}}

\begin{document}

\title{Does the coframe geometry can serve as  a unification scheme?}

\author{Yakov Itin}

\affiliation{Institute of Mathematics, Hebrew University of
  Jerusalem } 

\begin{abstract}
 The coframe field model is known as a viable model for gravity.
The principle problem is an interpretation of six additional
 degrees of freedom.
 We  construct a general family of connections which includes the
 connections of Levi-Civita and Weitzenb\"{o}ck as the limiting cases.
  We show that for a special choice of parameters,
 a subfamily of  connections is invariant
 when the infinitesimal field of transformations
 (antisymmetric tensor) satisfies the pair
 of vacuum Maxwell equations --- one for torsion and one for non-metricity. 
Moreover, the vacuum Maxwell equations turn to be the  
necessary and sufficient conditions for invariance of the  viable 
coframe action (alternative to GR). 
Consequently, for the viable models, the coframe field is proved to have the 
Maxwell-type behavior in addition to the known gravity sector. 

\end{abstract}

\date{\today}
\maketitle


\section{Introduction. Absolute frame variable}
 Absolute (teleparallel, fernparallel, ...) frame variable $e_\a$ was
 introduced in physics by Einstein in 1928 \cite{cit1}.
 This object is rather widely used in standard GR as well as
 in its various alternations, see \cite{cit1x} and the reference given
 therein.
 The frame  $e_\a$ and its dual, the coframe $\vt^\a$,
 have a well defined geometrical sense. In particular,
 they may be used as a reference basis.
 For a fixed absolute basis $\{e_\a\,,\vt^\a\}$,  this construction
 gives an invariant
 meaning to the components of a tensor, thus it emerges violation
 of Lorentz invariance.
 However, allowing global
 Lorentz transformations of the absolute frame field,
 the frame components of a tensor are merely
 transformed by the Lorentz transformation law.
 Thus, some interrelation emerges between Lorentz
 invariant field theories
 and diffeomorphism invariant gravity.

 In local coordinates \cite{cit2}, the frame field,
 $e_\a={e_\a}^{a}\,\partial_a$, and its dual,
 the coframe field, $\vt^\a={\vt^\a}_a \,dx^a$,
 are expressed  by $4\times 4$
 matrices
which are reciprocal  to each other
\begin{equation}\label{recip}
{e_\a}^{a}\,{\vt^\b}_a=\d^\b_\a\,, \qquad
{e_\a}^{a}\,{\vt^\a}_b=\d^a_b\,.
\end{equation}
Thus we have another intrigued property of the frame field
variable:
 $16=10+6$. It is most desirable to have a separation of
 sixteen independent
 variables of the frame field to ten variables for gravity plus six
 variables for electromagnetic field.
 Unfortunately, this idea does not work in such a
  simple form.
 Ten gravity variables are easily extracted from the coframe
 by use of the metric
 tensor ($\eta_{\a\b}= {\rm{diag}}(-1,1,1,1)$)
  \begin{equation}\label{metr}
g=\eta_{\a\b}\,\vt^\a\otimes \vt^\b\,, \qquad
 g_{ab}=\eta_{\a\b}\,\vt^\a{}_a\,\vt^\b{}_b\,.
\end{equation}
As for the six remaining components of the coframe,
 a corresponded Lorentz invariant algebraic combination
 fails to exist. Consequently, the problem is:

 What physical interpretation can be given to these six remained
degrees of freedom?
 More pretentiously, is it possible to  extract six
 electromagnetic field strengths   from the frame/coframe variables?

 \section{Coframe field model}
 Let us start with a brief account of the coframe field model,
 see \cite{Itin:2001bp} for an exterior form representation.
 Let a manifold be endowed with two smooth basis fields
 $\vt^\a(x)$ and $e_\a(x)$,
 which are assumed to be fixed up to global (rigid)
 Lorentz transformations.

 The action is required to be  quadratic in the first order derivatives
 $\vt^\a{}_{m,n}$.
 Although the second order derivative terms in the
 form of total divergence are also admissible, we neglect such
 additions.
 A global Lorentz and diffeomorphism  invariant action functional
 may be easily constructed from the  exterior derivative
 components $\vt^\a{}_{[b,c]}$.
 Since it is preferable to deal with a quantity who's indices
 are all of the same nature, we introduce two  tensors
 \begin{equation}\label{Weiz}
 \oT_{ab}{}^c=e_\a{}^c\,\vt^\a{}_{[a,b]}\,,\qquad
 C_{\a\b}{}^\g=e_\a{}^ae_\b{}^b\,\vt^\g{}_{[a,b]}\,.
 \end{equation}
 A general ``quadratic'' action functional may be
written now as
 \begin{equation}\label{act1}
\A=\frac{\kappa}2\int C_{\a\b\g} \F^{\a\b\g}\,\sqrt{-g}\,dx^4+
{}^{\rm (m)}\A\,.
\end{equation}
 Here $\kappa$ is a coupling constant, ${}^{\rm (m)}\A$ is an action for
 a matter field, and $\F^{\a\b}{}_\g$ is a tensor which is assumed to be
 linear in the components of $C_{\a\b}{}^\g$:
  \begin{equation}\label{F-def}
 \F^{\a\b\g}=\lambda^{\a\b\g\mu\nu\rho}\,C_{\mu\nu\rho}\,,
 \end{equation}
 The tensor $\lambda^{\a\b\g\mu\nu\rho}$ carries
 only  the frame indices, so its general expression may be written  as
 \begin{equation}
 \label{lambda}
\lambda^{\a\b\g\mu\nu\rho}=\mu_1\eta^{\a\mu}\eta^{\b\nu}\eta^{\g\rho}+
\mu_2\eta^{\a\nu}\eta^{\b\rho}\eta^{\g\mu}+
\mu_3\eta^{\a\g}\eta^{\mu\rho}\eta^{\b\nu}\,.
\end{equation}
 The following change of parameters
 \begin{equation}
 \label{rho-mu}
 \mu_1=\rho_1+\rho_2+\rho_3\,, \quad \mu_2=2\rho_2\,,\quad \mu_3=-2\rho_3
 \end{equation}
 appears to be useful for the exterior form representation \cite{Itin:2001bp}.
With (\ref{lambda}), the action functional (\ref{act1}) may be
 rewritten in a compact exterior differential form
 \begin{equation}\label{act2}
\A=\frac{\kappa}4\int d\vt^\a\wedge *\F_\a+{}^{\rm (m)}A\,.
\end{equation}
Here, $*$ denotes the Hodge dual while the 2-form  $\F^\a$ is
  \begin{equation}\label{F-str}
 \F^\a=\frac 12 \F^\a{}_{\b\g}\vt^\b\wedge\vt^\g=
 \frac 12 \F^\a{}_{\b\g}\vt^\b{}_m\vt^\g{}_n\,dx^m\wedge dx^n\,.
 \end{equation}
%
%
Variation of  (\ref{act2})  yields the coframe field
equation
  \begin{equation}\label{feq}
d*\F^\a={}^{\rm (c)}\T^\a+{}^{\rm (m)}\T^\a\,.
\end{equation}
Here ${}^{\rm (m)}\T^\a=\d({}^{\rm (m)}\A)/\d\vt^\a$
 is the 3-form of the matter energy-momentum
current.
 Such an object is related to the energy-momentum tensor as
 $\T^\a=\T^\a{}_\b*\vt^\b$.
 The additional 3-form ${}^{\rm (c)}\T^\a$
comes from  variation of  the Hodge dual operator which itself
depends on the coframe. Being related to the absolute frame, the
corresponded energy-momentum tensor takes
 the regular Yang-Mills form,
  \begin{equation}\label{en-mom}
 {}^{\rm (c)}\T^\mu{}_\nu=-\,C^{\a\mu\b}\F_{\a\nu\g}+
 \frac 14\, \d^\mu_\nu \,C^{\a\b\g}\F_{\a\b\g}\,.
\end{equation}
 Since the right hand side of  (\ref{feq}) is conserved,
 it has  to be identified as the total energy-momentum current
 for the system of the coframe and  matter fields.
 Consequently, ${}^{\rm (c)}\T^\a$ accepts the meaning of the
 coframe energy-momentum current. This Hilbert type consideration
 is supplemented  with the Noether procedure, which yields the same
 expression (\ref{en-mom}), see \cite{Itin:2001bp}.

For a generic choice of the parameters $\rho_i$, (\ref{feq}) is a
well posed system of 16 independent equations for 16 independent
coframe components. Being a vector-valued 3-form expression, the
field equation may be covariantly reduced to a system of two
tensorial equation, the symmetric equation and
 the antisymmetric one. We write them symbolically as
\begin{equation}\label{Eq-red}
{\cal EQ}_{(ab)}={}^{\rm (m)}T_{(ab)}\,,\qquad {\cal
EQ}_{[ab]}={}^{\rm (m)}T_{[ab]}\,.
\end{equation}
The  left hand side of the antisymmetric equation
 vanishes identically if and only if
  \begin{equation}\label{E-choice}
\rho_1=0\,,\qquad \rho_2=- 1/2 \,,\qquad \rho_3=1.
\end{equation}
In this case, we remain with a system of only 10 independent
equation. So only 10 combinations of the coframe are determined.
Although, it is enough to determine the metric tensor (\ref{metr}).

The coframe energy-momentum tensor (\ref{en-mom}) is well defined
for a generic set of parameters $\rho_i$. This object is invariant
under global Lorentz transformations and covariant under smooth
changes of coordinates. Moreover, the tensor (\ref{en-mom}) is
traceless, in accordance to the scale invariance of the coframe
Lagrangian.
 \section{Gravity sector}
 Let us briefly recall how the coframe field model works in the
 gravity sector.
Observe that every  polynomial constructed from the derivatives of
the metric is expressed, by (\ref{metr}), in the derivatives of the
coframe. Consequently, the standard Einstein-Hilbert action has to
appear as a special case of the general coframe action. Indeed,
the action (\ref{act2}) with the parameters (\ref{E-choice}) is
equivalent to the Einstein-Hilbert action (up to a total
derivative).
 For this set of parameters,  the action
(\ref{act1}) and the field equation (\ref{feq}) have a hidden
symmetry: They are invariant under local Lorentz  transformations,
\begin{equation}\label{trans}
\vt^\a\mapsto L^\a{}_\b \vt^\b\,, \qquad L^\a{}_\b(x)\in SO(1,3)\,.
\end{equation}

For the parameters (\ref{E-choice}), the  equation
(\ref{feq}) must have a coframe solution corresponded by
(\ref{metr}) to the Schwarzschild metric. It is natural
 to look for possible spherical-symmetric solutions of (\ref{feq}) with an
arbitrary set of parameters.
The answer is as following \cite{Itin:1999wi}:

 (i) For the set of the parameters
 \begin{equation}\label{arb-par}
\rho_1=0\,,\qquad \rho_2={\rm arbitrary} \,,\qquad \rho_3=1\,,
\end{equation}
the field equation (\ref{feq}) has a unique static
spherical-symmetric solution of a "diagonal form":
\begin{equation}\label{Sch}
\vt^0=\frac{1-m/2r}{1+m/2r}\,dx^0\,,\qquad \vt^i=
\left(1+\frac{m}{2r}\right)^2\,dx^i\,,
\end{equation}
where $i=1,2,3$. This coframe corresponds to the Schwarzschild metric
 in the isotropic coordinates.

 (ii) If the parameters differ from (\ref{arb-par}), any exact
solution of a "diagonal form" does not have the Newtonian behavior
at infinity.

Thus, the condition $\rho_1=0$ identifies a family of viable models
with Schwarzschild solution. Another justification of this
condition comes from consideration of the first order approximation
to the coframe field model \cite{Itin:2004ig}. In this case, the coframe
variable is reduced to a sum of symmetric and antisymmetric
matrices. It means that, in linear approximation,
 we can treat the coframe field as a system
of two independent fields. It is natural to require all the
field-theoretic constructions, i.e. the action, the field equation
and the energy-momentum tensor, to accept the same reduction to
two independent expression. It is   remarkably  that such
separation appears if and only if $\rho_1=0$.

Thus, instead of a unique gravity model based on the Riemannian
metric, for the coframe variable, we have a whole family
(\ref{arb-par}) of viable gravity models parameterized by the
parameter $\rho_2$.

For $\rho_2=-1/2$, the coframe model is equivalent to
the standard GR. The equivalence proved on the level of the action,
of the field equation and of the exact solutions. The
energy-momentum expression (\ref{en-mom}), however, is not
invariant  under  local transformations of the coframe.
Actually,  it is no more than a type of a pseudo-tensor.  The local
invariance also decreases the number of degrees of freedom to ten
metric components.

For $\rho_2\ne-1/2$, we have an alternative model  of
sixteen independent degrees of freedom. The action is only global
Lorentz invariant so it is not correct to require more symmetries
 for the energy-momentum tensor.
 In particular, the field equation and the energy-momentum
tensor of the coframe field are well defined. So the
energy-momentum problem of GR is solved in this alternative
context. The price is a set of new problems: (i) What
interpretation can be given for the additional six degrees of
freedom? (ii) Which value of the parameter $\rho_2$ must be
chosen? (iii) What geometries can be related to different values of
the parameters?
 \section{``Geometrization'' of the coframe model}
 Although the coframe variable itself has a well defined geometrical sense,
 the action (\ref{act1}) and the field equation (\ref{feq}) are not
 related yet to any specific geometry.

 We accept the Cartan viewpoint which treats a
 geometrical structure as a pair $\{g_{ab}\,,\G_{ab}{}^c\}$ of two
 independent objects:
 the metric field $g_{ab}$ and the connection field $\G_{ab}{}^c$.
 Moreover, we require $g_{ab}$ and $\G_{ab}{}^c$ to be explicitly
 constructed from the coframe components.

 The metric tensor (\ref{metr}) is already constructed from the coframe.
 Due to the index content, this construction is  unique
 (up to a scalar factor).

 As for the field of connection $\G_{ab}{}^c$, we require
 it to be linear in the first order derivatives of the coframe
 components $\vt^\a{}_{m,n}$.
 The coefficients in this linear combination are polynomial
 in the coframe components.
 Observe that the similar  requirement, being accepted in the Riemannian
 geometry when the connection is linear in the first order
 derivatives of the metric, gives a unique Levi-Civita connection.
 In the coframe background the situation is rather different: In fact,
 we have here a whole family of connections. Recall the properties:
 (i) The connection is a set of $4^3$
 components which change by a specific inhomogeneous linear law;
 (ii) The difference of two connections is a tensor of type
 $(1,2)$.

 On a  manifold  endowed with a coframe field an absolute
 (curve independent)
 sense can be given to  the parallelism of distance vectors.
 Namely, two vectors may be declared parallel one to another when
 they have the proportional components being referred to the
 local absolute frames.
 It means that the covariant  derivatives  of the absolute
 coframe components are zero relative to some special connection
 $\oG_{bc}{}^a$, which is referred to as the Weitzenb\"{o}ck connection.
 From ${\vt^\a}_{a;b}=0$, 
 we have, by  (\ref{recip}), the  Weitzenb\"{o}ck connection as
  \begin{equation}\label{weiz2}
 \oG_{ab}{}^c={e_\a}^c\,{\vt^\a}_{a,b}\,.
 \end{equation}
 Under a transform of coordinates, this expression changes by
 the proper inhomogeneous linear law.

 A general coframe connection may be  represented now
as the Weitzenb\"{o}ck connection plus a
 tensor of type $(1,2)$
\begin{equation}\label{con}
\G_{ab}{}^c=\oG_{ab}{}^c+K_{ab}{}^c\,,
\end{equation}
where $K_{bc}{}^a$  is linear in the first order derivatives
\begin{equation}\label{add-tens}
K_{ab}{}^c=\chi_{abm}{}^{cnp}\vt^\a_{[n,p]}e_\a{}^m=
 \chi_{abm}{}^{cnp}\oG_{[np]}{}^m\,.
\end{equation}
The general form of the "coupling tensor" is 
 \begin{eqnarray}\label{mcc8}
&&\chi_{abm}{}^{cnp}=\a_1\d_m^c\d_a^{n}\d_b^{p}+
\a_2\d_a^c\d_m^{n}\d_b^{p}+\a_3\d_b^c\d_m^{n}\d_a^{p}\nonumber\\
&&+\b_1g_{ab}g^{cn}\d^{p}_m+\b_2g_{am}g^{cn}\d^{p}_b+
\b_3g_{bm}g^{cn}\d^{p}_a\,.
\end{eqnarray}
 Consequently the additional ("contortion") tensor is
\begin{eqnarray}\label{K-tens}
 &&\!\!\!\!\!\!K_{ab}{}^c=\a_1\oT_{ab}{}^c+\a_2\d_a^c\oT_{mb}{}^m+
 \a_3\d^c_b\oT_{ma}{}^m+\nonumber\\
&&\!\!\!\!\!\!g^{cn}\left(\b_1g_{ab}\oT_{nm}{}^m+\b_2g_{am}\oT_{nb}{}^m+
\b_3g_{bm}\oT_{na}{}^m\right)\,.
\end{eqnarray}
Hence, in contrast to the metric geometry, we have  a 6-parametric
connection constructed from the coframe components. Every
connection $\G_{bc}{}^a$ is characterized by two tensors. The
torsion, $T_{ab}{}^c=-T_{ba}{}^c$, is defined as
\begin{equation}\label{tor}
T_{ab}{}^c=\G_{[ab]}{}^c\,.
\end{equation}
The non-metricity tensor, $Q_{cab}=Q_{cba}$, is  defined as
\begin{equation}\label{nonmetr}
Q_{cab}=-\nabla_cg_{ab}=
 -g_{ab,c}+\G_{acb}+\G_{bca}\,.
\end{equation}
where $\G_{acb}=\G_{ac}{}^mg_{mb}$.

For the  Weitzenb\"{o}ck connection, the torsion is given by the
 $\oT_{bc}{}^a$ while the non-metricity tensor is zero.
The Levi-Civita connection is defined by setting  both tensors to
zero.

For the general connection (\ref{con},\ref{K-tens}), the torsion is
 \begin{eqnarray}\label{g-tor}
T_{ab}{}^c&=&(1+\a_1)\oT_{ab}{}^c+\nonumber\\&&\frac 12
(\a_2-\a_3)(\d^c_a\oT_{mb}{}^m-\d^c_b\oT_{ma}{}^m)+\nonumber\\
 &&\frac 12(\b_2-\b_3)g^{cn}(g_{am}\oT_{nb}{}^m-g_{bm}\oT_{na}{}^m)\,.
\end{eqnarray}
 Consequently, the  connection (\ref{con}) is identically
 torsion-free if and only if
\begin{equation}\label{tor-free}
 \a_1=-1\,,\qquad \a_3=\a_2\,,\qquad \b_2=\b_3\,.
\end{equation}
The non-metricity tensor of the general connection
 (\ref{con},\ref{K-tens}) is
 \begin{eqnarray}\label{g-nonmetr}
Q_{cab}&=&(\a_1+\b_2)(\oT_{acb}+\oT_{bca})+2\a_2g_{ab}\oT_{mc}{}^m+
\nonumber\\
 &&(\a_3-\b_1)(g_{ac}\oT_{mb}{}^m+g_{bc}\oT_{ma}{}^m) \,.
\end{eqnarray}
Hence, the  connection (\ref{con}) is metric-compatible
  if
  \begin{equation}\label{mcc43}
 \a_1=-\b_2\,, \qquad \a_2=0\,, \qquad \a_3=\b_1\,.
 \end{equation}
Thus, the unique torsion-free and metric-compatible coframe
connection is given by the set of parameters
\begin{equation}\label{mcc44}
 -\a_1=\b_2=\b_3=1\,, \qquad \a_2=\a_3=\b_1=0\,.
 \end{equation}
 Certainly, it is not more than the ordinary Levi-Civita connection
 (\ref{LC-con}),
 which we can  express now by the Weitzenb\"{o}ck
connection
\begin{equation}\label{mcc45}
 \LG_{abc}=\oG_{(ab)c}+\oG_{[ca]b}+\oG_{[cb]a}\,.
 \end{equation}
 Two tensors, the torsion an the non-metricity,  characterize the
 coframe connection uniquely. 
 Indeed, let a metric $g$  be fixed and two
tensors $T_{ab}{}^c=-T_{ba}{}^c$ and $Q_{cab}=Q_{cba}$
 be given. The corresponding unique connection is \cite{Schou}.
\begin{eqnarray}\label{LC-decomp}
\G_{abc}&=&\LG_{abc}+(T_{abc}+T_{cab}-T_{bca})+\nonumber\\
 && \frac 12 (Q_{abc}-Q_{cab}+Q_{bca})\,,
\end{eqnarray}
where $\LG_{abc}$ is used for the Levi-Civita connection:
\begin{equation}\label{LC-con}
 \LG_{abc}=\frac 12 (g_{ac,b}+g_{bc,a}-g_{ab,c})\,
 \end{equation}
\section{Maxwell equations}
The connection (\ref{con}) is invariant under global (rigid)
transformation of the absolute coframe. We will now examine how it
changes under local linear transformations $\vt^\a\mapsto
L^\a{}_\b(x)\,\vt^\b$. In a coordinate basis,   the coframe
components change as
 \begin{equation}\label{L-tr1}
 \vt^\a{}_a\mapsto L^\a{}_\b\,\vt^\b{}_a\,,\qquad
 e_\a{}^a\mapsto (L^{-1})^\b{}_\a\, e_\b{}^a\,.
\end{equation}
The infinitesimal version of this transformation with
$L^\a{}_\b=\d^\a_\b+X^\a{}_\b$ is
 \begin{equation}\label{L-tr2}
 \vt^\a{}_a\mapsto \vt^\a{}_a+X^\a{}_\b\,\vt^\b{}_a\,,\qquad
 e_\a{}^a\mapsto e_\a{}^a-X^\b{}_\a\, e_\b{}^a\,.
\end{equation}
 Let us examine now, under what conditions the geometrical structure
 is invariant under these transformations.
 Since the coframe field appears in the geometrical structure,
 $\{g_{ab}(\vt^\a)\,,\G_{ab}{}^c(\vt^\a)\}$, only implicitly, (\ref{L-tr1})
 is a type of a gauge transformation.
 Invariance of the metric tensor restricts $L^\a{}_\b$ to
 a pseudo-orthonormal matrix.
  In the infinitesimal version,
 it means that the matrix $X_{\a\b}=X^\mu{}_\b \eta_{\mu\a}$
 is antisymmetric.

 The Levi-Civita connection $\LG_{bc}{}^a$ is  invariant under the
 transformations (\ref{L-tr1}) with an arbitrary matrix $X_{ab}$.
 As for the Weitzenb\"{o}ck connection, it changes as
 \begin{equation}\label{W-change}
 \Delta \oG_{abc}=\vt^\a{}_c\vt^\b{}_aX_{\a\b,b}\,.
 \end{equation}
 Thus the Weitzenb\"{o}ck connection is invariant only if $X^\a{}_{\b,a}=0$,
 i.e., only for rigid transformations.

 Let us ask now, under what conditions the  connection {\ref{con})
 of a nonzero torsion and non-metricity  is invariant.
 We apply the decomposition (\ref{LC-decomp}).
 Since $\LG_{bc}{}^a$ is invariant,
 we have to require two tensorial equations:
 \begin{equation}\label{TQ-change}
  \Delta T_{abc}=0\,, \qquad  \Delta Q_{abc}=0\,.
  \end{equation}
  For a generic set of parameters $\a_i,\b_i$, the left hand
  sides of (\ref{TQ-change}) do not vanish identically.
  Instead,  we have here two first order
  partial differential equations
  for the antisymmetric tensor field $X_{\a\b}$.
 Instead of $X_{\a\b}$ we define an
 antisymmetric matrix with coordinate indices
 \begin{equation}\label{F1-def}
 F_{ab}=X_{\mu\nu}\vt^\mu{}_a\vt^\nu{}_b\,.
 \end{equation}
 Moreover let us restrict to the case when
 the derivatives  of the coframe field are small comparing to the
 derivatives of the matrix  $X_{\mu\nu}$.
 Under this condition, (\ref{W-change}) may be rewritten as
  \begin{equation}\label{con-change}
 \Delta \oG_{abc}=F_{ca,b}\,.
 \end{equation}
 Under the transformations (\ref{L-tr2}),
 the torsion (\ref{g-tor}) changes as
 \begin{eqnarray}\label{tor-change}
 \Delta T_{abc}&=&\frac 12 (1+\a_1)(F_{ca,b}-F_{cb,a})-\nonumber\\
 &&
 \frac 14 (\a_2-\a_3)(g_{ac}F^m{}_{b,m}-g_{bc}F^m{}_{a,m})+\nonumber\\
 &&\frac 14 (\b_2-\b_3)(F_{ac,b}-2F_{ab,c}-F_{bc,a})\,.
 \end{eqnarray}
 For  a special family of connections with parameters
\begin{equation}\label{scon}
\a_2=\a_3\,,\qquad (\b_2-\b_3)+2(1+\a_1)=0\,,
  \end{equation}
 we obtain
 \begin{equation}\label{tor-change1}
 \Delta T_{abc}=(1+\a_1)(F_{ab,c}+F_{bc,a}+F_{ca,b})\,.
  \end{equation}
 Under the transformations (\ref{L-tr2}),
 the non-metricity  (\ref{g-nonmetr}) changes as
 \begin{eqnarray}\label{nonmetr-ch}
 \Delta Q_{cab}&=&-\frac 12 (\a_1+\b_2)(F_{bc,a}+F_{ac,b})-
 \a_2g_{ab}F^m{}_{c,m}\nonumber\\
 &&-\frac 12 (\a_3-\b_1)(g_{ac}F^m{}_{b,m}+g_{bc}F^m{}_{a,m})\,.
 \end{eqnarray}
 For  a family of connections with parameters
\begin{equation}\label{scon1}
\a_1+\b_2=0\,, \qquad \a_3-\b_1=0\,,
  \end{equation}
 we have
 \begin{equation}\label{nonmetr1-ch}
 \Delta Q_{cab}=-\a_2g_{ab}F^m{}_{c,m}
  \end{equation}
 Consequently, we have derived a nonempty family of connections
 that are invariant under the
 transformation (\ref{L-tr2}) provided the antisymmetric tensor $F_{ab}$
 satisfies the Maxwell  field equations
  \begin{equation}\label{max}
 F_{ab,c}+F_{bc,a}+F_{ca,b}=0\,,\qquad F^m{}_{a,m}=0\,.
 \end{equation}

 
Let us look now how these ``Maxwell transformations'' are connected to 
the coframe Lagrangian. Recall that we are looking for a symmetry which 
distinguishes the viable models with Schwarzschild solutions.  
Since the second order term is involved only as a total derivative, 
we can write the viable Lagrangian as 
 \begin{eqnarray}\label{vc2}
{}^{(v)}L&=&R+\frac 12(2\rho_2+1)\oT^{abc}(\oT_{abc}+2\oT_{cab})+\nonumber \\&&
2(\rho_3-1)\oT_m{}^{am}\oT^n{}_{an}
 \end{eqnarray}
Under the transformations (\ref{L-tr2}), 
this Lagrangian is transformed as 
  \begin{eqnarray}\label{vc3}
\Delta \Big({}^{(v)}L\Big)&=&\frac 12(2\rho_2+1)\oT^{abc}(F_{ab,c}+F_{bc,a}+F_{ca,b})+\nonumber\\&&
2(\rho_3-1)\oT_m{}^{am}F^n{}_{a,n}
\end{eqnarray}
Thus, for the non-Einstein models, the Lagrangian is invariant if and only if the local Lorentz transformations satisfy the vacuum 
Maxwell equations (\ref{max}). 

\section{Discussion}
We discuss briefly and somewhat speculatively how this construction
can work.\\
\noindent
 1. {\it Variables.} The set of coframe fields is separated to
 equivalence classes,
 while the equivalence relation is given by (infinitesimal)
 Lorentz transformations.
 The fields from the same equivalence class generate the same
 metric which is
 a representation of the gravity field. The antisymmetric
 tensor of infinitesimal
 Lorentz transformations represents the electromagnetic field.
\\
\noindent
 2. {\it Action.} An action for a system of a metric field and a field
 of connection is known
 from the metric-affine gravity \cite{Hehl:1994ue}.
 It is represented as a  sum of 28 terms --- the squares of irreducible
 pieces of torsion and non-metricity plus the irreducible pieces
 of the curvature.
 In the coframe framework, the metric and the connection are
 constructed from the derivatives of the coframe components.
 Consequently, in this model,  the whole metric-affine
 action is no more then the action (\ref{act1}) (up to a total derivative).
\\
\noindent
3. {\it Field equations.}
In fact, the equations (\ref{max}) mean certain constrains on the 
 coframe field. But the field equations (\ref{feq}) compose a well posed 
 system. Two facts are  coordinate one to another only if some six equations 
 from (\ref{feq}) are the consequence of the Maxwell equations. 
 It is true at least for the linear approximation of (\ref{feq}), 
 see \cite{Itin:2004ig}. In this case, (\ref{feq}) is reduced to two 
 independent systems. 
 One of them is merely $\square F_{ab}=0$, i.e., a 
 consequence of  (\ref{max}). 
\\
\noindent
4. {\it Sources.} We considered above the vacuum case only. 
 When the sources are modeled as singularities of the  fields,  
 we can also involve singular Lorentz transformations. 
 Since the transformations play the role of the equivalent relations, it is 
 natural to require them not to create new singularities in addition 
 to the singularities of the coframe field.   
 Thus the Lorentz transformations can be of two types: (i) Nonsingular, 
 (ii) Singular in the same point as the coframe is. 
 A remarkable physical consequence of this consideration is:   
 The massive points can be charged and uncharged while the charged points 
 must be massive. Also the absence of the Dirac monopole may be 
 related to the absence of the massive string solutions. \\
\noindent 5. {\it Geodesics.}
The motion of singularities is described by the geodesic equation, 
 which does not involve the torsion part of the connection. 
 Thus, in order to have the Lorentz force in an addition 
 to the Newtonian one,  the connection has to contain the non-metricity 
 ingredient.   
\section*{Acknowledgment} 
I thank Friedrich Hehl and Roman Jackiw for valuable discussion. 


\begin{thebibliography}{99}
\bibitem{cit2} The Roman index refers to a specific coordinate,
 while the Greek index  is a label for the absolute basis element
 involved. These  indices are principally
 different, in particular, they cannot be contracted in  ${\vt^\a}_a$
 or ${e_\a}^{a}$.
\end{thebibliography}
\end{document}